\documentclass[useAMS,usenatbib]{mn2e}

\usepackage[pdftex]{graphicx}
\usepackage[fleqn]{amsmath}
\usepackage[usenames]{color}

\newcommand{\rp}{r_{\mathrm p}}

\newcommand{\op}{\Omega_\mathrm{p}}
\newcommand{\okep}{\Omega_\mathrm{K}}
\newcommand{\cs}{c_\mathrm{s}}

\newcommand{\xs}{x_\mathrm{s}}
\newcommand{\zs}{z_\mathrm{s}}

\newcommand{\be}{ \begin {equation}}
\newcommand{\ee}{ \end {equation}}

\newcommand{\sigp}{\Sigma_\mathrm{p}}
\newcommand{\nup}{\nu_\mathrm{p}}
\newcommand{\chip}{\chi_\mathrm{p}}
\newcommand{\Fp}{f_\mathrm{p}}
\newcommand{\yp}{y_\mathrm{p}}
\newcommand{\tturn}{\tau_\mathrm{turn}}
\newcommand{\tdiff}{\tau_\mathrm{diff}}

\title[Non-isothermal Type I planetary migration -II]{A torque formula for non-isothermal Type I planetary migration - II. Effects of diffusion}
\author[S.-J. Paardekooper, C. Baruteau and W. Kley]{S.-J. Paardekooper$^{1}$\thanks{E-mail:
S.Paardekooper@damtp.cam.ac.uk}, C. Baruteau$^{2}$ and W. Kley$^{3}$\\
$^{1}$DAMTP, University of Cambridge, Wilberforce Road, Cambridge CB3 0WA, United Kingdom\\
$^{2}$UCO/Lick Observatory, UC Santa Cruz, 1156 High Street, Santa Cruz, CA 95064, USA\\
$^{3}$Institut f\"ur Astronomie \& Astrophysik, Universit\"at T\"ubingen, Auf der Morgenstelle 10, 72076 T\"ubingen, Germany}

\begin{document}

\date{Draft version \today}

\pagerange{\pageref{firstpage}--\pageref{lastpage}} \pubyear{2008}

\maketitle

\label{firstpage}

\begin{abstract}
We study the effects of diffusion on the non-linear corotation torque, or horseshoe drag, in the two-dimensional limit, focusing on low-mass planets for which the width of the horseshoe region is much smaller than the scale height of the disc. In the absence of diffusion, the non-linear corotation torque saturates, leaving only the Lindblad torque. Diffusion of heat and momentum can act to sustain the corotation torque. In the limit of very strong diffusion, the linear corotation torque is recovered. For the case of thermal diffusion, this limit corresponds to having a locally isothermal equation of state. We present some simple models that are able to capture the dependence of the torque on diffusive processes to within $20$ $\%$ of the numerical simulations.
\end{abstract}



\section{Introduction}
Low-mass planets, embedded in circumstellar gas discs, are subject to orbital migration \citep{GT79,GT80} through disc tides. Until they can acquire a massive gas envelope, planets comparable in mass to the Earth can therefore significantly move away from their place of birth. It was long thought that this Type I migration was always directed inward, and alarmingly fast \citep{ward97,tanaka02}, which suggested that the survival of long-period planets could be problematic within this theory. Recent work has indicated that the situation may be different if the disc's thermodynamics is taken into account properly \citep{paard06,kley08,kley09}, in which case outward migration is possible. It was subsequently realised that this migration behaviour is due to an entropy-related corotation torque \citep{paard08,baruteau08}, and is non-linear in nature \citep{paardpap08}.   

In \cite{paard10}, hereafter Paper I, we studied the non-linear corotation torque, or horseshoe drag \citep{ward91}, in the two-dimensional adiabatic limit. It was shown that this horseshoe drag is dominated by a term proportional to a radial entropy gradient in the unperturbed disc, and can lead to outward migration if the entropy decreases outward. In the absence of any diffusion, however, the corotation torque is a transient phenomenon and prone to saturation \citep{quinn86,ward07}. The necessary gradients (vortensity in the (locally) isothermal limit, entropy and vortensity for the non-isothermal case) within the horseshoe region need to be restored on a libration time scale in order for the corotation torque to be sustained \citep{mas01,ogilvie03}. It was shown in \cite{paardpap08} that for non-isothermal discs, both thermal diffusion and viscosity are needed to sustain the non-linear corotation torque. 

Apart from possibly keeping the corotation torque unsaturated, diffusion of heat and momentum can also, in the limit of high diffusion, act to reduce the non-linearity of the torque. It was already noted in \cite{mas02} that there exists a cut-off for the isothermal horseshoe drag at high viscosity. \cite{drag} showed that in this limit, the corotation torque returns to its linear value. For non-barotropic discs, the limit of high thermal diffusion effectively corresponds to the locally isothermal case. 

Depending on radial profiles of density, temperature, viscosity and thermal diffusion, the latter mainly determined by temperature and opacity, a wealth of Type I migration behaviour is possible therefore. While the sign of the unsaturated torque is independent of planet mass, saturation does depend on the mass of the planet, which can lead to different migration directions for different planet masses. Moreover, since the level of saturation effectively determines whether the non linear corotation torque can dominate Type I migration, it is important to study the effects of diffusion on low-mass planet migration. 
    
In this paper, we study Type I migration in non-barotropic discs in the presence of thermal and viscous diffusion. We start in section \ref{secNum} by reviewing our disc models and numerical methods. In section \ref{secUnsat}, we briefly review the results of Paper I on the unsaturated values of the torque. We then move on to the effects of viscous diffusion in section \ref{secIso} by considering the simple isothermal case, first looking at saturation, and then moving to the limit of high diffusion in which the linear torque is recovered. We extend these results to the non-barotropic case in section \ref{secNonbaro}. We discuss our results in section \ref{secDisc} and conclude in section \ref{secCon}.

\section{Numerical Methods}
\label{secNum}
For the basic equations and details of the equilibrium disc models we refer the reader to Paper I. As in Paper I, we study two-dimensional power law discs in surface density $\Sigma$ (with index $-\alpha$) and temperature $T$ (with index $-\beta$). We study either isothermal discs, for which the pressure $P=\cs^2\Sigma$, with $\cs$ the isothermal sound speed, or solve the energy equation and take a perfect gas equation of state, $P=(\gamma-1)\epsilon$, where $\epsilon$ is the internal energy density and $\gamma$ the adiabatic exponent. The scale height of the disc $H=\cs/\Omega$ is taken to be small compared to the orbital radius. Typically $h=H/r=0.05$ at the location of the planet.

We have used the same numerical methods as in Paper I. {\sc fargo}  (Fast Advection in Rotating Gaseous Objects, \citet{fargo,fargo2}) is based on the van Leer upwind algorithm on a staggered mesh. The time step calculation takes out the average angular velocity, which is dominated by Keplerian rotation, so that the time step is limited by the residual angular velocity rather than the much larger Keplerian velocity. An energy equation solver was introduced in \cite{baruteau08}. Diffusion in the energy equation is obtained by adding a term proportional to the Laplacian of the entropy:
\begin{equation}
\frac{\partial \epsilon}{\partial t}+\nabla\cdot \epsilon {\bf v}=-(\gamma-1)\epsilon \nabla \cdot {\bf v}+\chi \epsilon \nabla^2 \log \frac{P}{\Sigma^\gamma},
\end{equation}
where ${\bf v}$ is the velocity and $\chi$ a thermal diffusion coefficient. This equation is strictly equivalent to an advection-diffusion equation for the gas entropy $p/\Sigma^\gamma$.  Using this form of the energy equation is a common approach in studies of turbulent energy transport \citep[see][]{shakura}. We take $\chi$ to be a constant throughout the disc.

{\sc rodeo} (ROe solver for Disc Embedded Objects, \citet{rodeo}) is a finite volume method, using an approximate Riemann solver \citep{roe} to calculate fluxes between computational cells. Both codes were shown to yield similar values for the unsaturated horseshoe drag in the adiabatic case (see Paper I), with {\sc rodeo} usually producing slightly larger values (up to 10\%). In {\sc rodeo}, explicit heat diffusion is implemented by solving
\begin{equation}
\frac{\partial T}{\partial t}=-\nabla\cdot (\chi \nabla T).
\label{eqHeatDiff}
\end{equation}
For {\sc rodeo}, we choose $\chi$ to be a function of radius such that the initial temperature profile is a stationary solution. We have found that this only becomes important when the diffusion time scale over the horseshoe width becomes comparable to the dynamical time scale. Equation \ref{eqHeatDiff} is formally only correct for constant density, but therefore has the advantage of affecting the temperature only, while viscosity affects only density structures. This way, thermal and viscous diffusion can be treated separately.

The two different approaches for obtaining diffusion were found to yield similar results, with differences less than 10\%. In both cases, the diffusion equations are solved explicitly, which leads to a restriction of the time step. Note that we have not included viscous dissipation in the equations: we assume that the background temperature profile is set by an equilibrium between viscous heating and radiative cooling \citep{kley08}, and that viscous and thermal diffusion will act to restore the original profile.

We take the viscosity $\nu$ to be a power law in radius such that the initial surface density profile is a stationary solution. It is easy to verify that this requires $\nu \propto r^{\alpha-1/2}$.

\section{Unsaturated torques}
\label{secUnsat}
Throughout this paper, we will work with a fixed gravitational smoothing length for the planet's potential, $b/h=0.4$.  We discuss possible effects of different smoothing lengths in section \ref{secForm}. In Paper I, we found the following unsaturated torques for this smoothing length, which we repeat here for convenience: the Lindblad torque
\begin{equation}
\gamma \Gamma_\mathrm{L}/\Gamma_0=-2.5-1.7\beta+0.1\alpha,
\label{eqLind}
\end{equation}
the barotropic part of the horseshoe drag
\begin{equation}
\gamma \Gamma_\mathrm{hs,baro}/\Gamma_0=1.1\left(\frac{3}{2}-\alpha\right),
\label{eqHSbaro}
\end{equation}
and the entropy-related part of the horseshoe drag
\begin{equation}
\gamma \Gamma_\mathrm{hs,ent}/\Gamma_0=7.9\frac{\xi}{\gamma},
\label{eqHSent}
\end{equation}
where $\xi=\beta-(\gamma-1)\alpha$ is the negative of the power law index of the entropy. The barotropic part of the linear corotation torque reads
\begin{equation}
\gamma \Gamma_\mathrm{c,lin,baro}/\Gamma_0=0.7\left(\frac{3}{2}-\alpha\right),
\label{eqCbaro}
\end{equation}
and the entropy-related part of the linear corotation torque is given by
\begin{equation}
\gamma \Gamma_\mathrm{c,lin,ent}/\Gamma_0=\left(2.2-\frac{1.4}{\gamma}\right)\xi.
\label{eqCent}
\end{equation}
As in Paper I, all torques are normalised to $\Gamma_0=(q/h)^2\Sigma_\mathrm{p}\rp^4\op^2$,  where subscripts $X_\mathrm{p}$ indicates quantity $X$ evaluated at the orbital radius of the planet $r=\rp$.

\section{Isothermal discs}
\label{secIso}

\subsection{A simple saturation model}
\label{secSimple}
\cite{mas01} discussed a model for corotation torque saturation. There are a few drawbacks to this approach, however, the major one being that the dependence of the corotation torque on background gradients has to be put in by hand. This actually leads to three possible saturation laws. Another difficulty is that since \cite{mas01} considers a constant viscosity $\nu$ and a constant background surface density, the unperturbed model has a radial velocity profile (i.e. it is not in equilibrium). At large viscosities, the coorbital region for low-mass planets can be deformed, resulting in a Type III flow structure. Below, we present a version of the \cite{mas01} model that is tailored to the problem at hand.

Assuming Keplerian rotation everywhere, the evolution of the surface density is given by
\begin{eqnarray}
\frac{\partial\Sigma}{\partial t}= \frac{3}{r}\frac{\partial}{\partial r}\left(\frac{1}{r\Omega} \frac{\partial}{\partial r}\left(r^2\nu\Sigma\Omega\right)\right)+\nonumber \\
\frac{|\Omega-\op|}{2\pi}\left(\Sigma'\frac{\Omega}{\Omega'}-\Sigma\right)\Pi\left(\frac{r-\rp}{2\rp\xs}\right),
\label{eqSigma}
\end{eqnarray}
where the second term on the right hand side accounts for changes due to the horseshoe turns \citep[see][]{mas01}. Primed quantities indicate they should be evaluated at $r'=2\rp-r$, i.e. on the other side of the horseshoe turn, and $\Pi$ denotes the rectangular function:
\begin{equation}
\Pi(t)=\left\{\begin{array}{rl}
1 & |t|< \frac{1}{2}\\
0 & |t| \geq \frac{1}{2}. \end{array}\right.
\end{equation}
The use of the rectangular function ensures that changes due to horseshoe turns only occur for $|r-\rp|/\rp<\xs$, where $\xs$ is the non-dimensional half-width of the horseshoe region. Changing variables to $y=\Sigma/\Omega$ and $x=(r-\rp)/\rp$, we have
\begin{eqnarray}
\lefteqn{\frac{\partial y(x,t)}{\partial t}=}\nonumber \\
& & \frac{3\nup}{\rp^2}(1+x)^{\frac{1}{2}}\frac{\partial}{\partial x}\left((1+x)^{\frac{1}{2}}\frac{\partial}{\partial x}\left( (1+x)^{\alpha-\frac{3}{2}}y(x,t)\right)\right)+\nonumber\\
& & \frac{|\Omega-\op|}{2\pi}(y(-x,t)-y(x,t))\Pi\left(\frac{x}{2\xs}\right).
\label{eqy}
\end{eqnarray}
Note that $y$ is proportional to the inverse of the specific vorticity, or vortensity, under the approximation of $\Omega$ being strictly the Keplerian angular velocity everywhere. Equation \ref{eqy} then shows that the vortensity changes with time due to diffusion (first term on the right-hand side) and due to advection in the horseshoe region (second term on the right-hand side). 

From now on, we will look for stationary solutions. Changing variables for a second time, with $z=2(1+x)^{1/2}-2$, and $f=(1+z/2)^{2\alpha-3}y(z)$, we find
\begin{eqnarray}
\lefteqn{\frac{3\nup}{\rp^2}\frac{\partial^2 f(z)}{\partial z^2}+} \nonumber \\
& & \frac{3\op}{4\pi}|z| \left[\left(1-\left(\frac{3}{2}-\alpha\right)z\right)f(-z)-\right.  \nonumber \\
& & \left. \left(1+\left(\frac{3}{2}-\alpha\right)z\right)f(z)\right]\Pi\left(\frac{z}{2\zs}\right)=0,
\end{eqnarray}
where we have approximated the second term of equation \ref{eqy} for $x\ll 1$ (or, equivalently, $z\ll 1$).  Here, $\zs=z(\xs)$. Following \cite{mas01}, we split $f$ into an even part, $\bar f$, and an odd part, $\tilde f$,  given by
\begin{equation}
\bar f(z) = \frac{1}{2}\left(f(z)+f(-z)\right),
\end{equation}
and
\begin{equation}
\tilde f(z) = \frac{1}{2}\left(f(z)-f(-z)\right).
\end{equation}
We then obtain finally
\begin{eqnarray}
\lefteqn{\frac{\partial^2 f(z)}{\partial z^2}-}\nonumber \\
& & \frac{\rp^2\op}{2\pi\nup}|z|\left(\tilde f(z)+\left(\frac{3}{2}-\alpha\right)z\bar f(z)\right)\Pi\left(\frac{z}{2\zs}\right)=0.
\label{eqf}
\end{eqnarray}
The even part of equation \ref{eqf} gives $\bar f=\Fp$, if we demand that $\bar f$ is bounded for all $z$. Note that $f=\Fp$ exactly corresponds to the unperturbed profile of $y(x)$.  Defining
\begin{equation}
k=\frac{\rp^2\op}{2\pi\nup},
\end{equation}
the odd part of equation \ref{eqf} gives
\begin{equation}
\frac{\partial^2 \tilde f(z)}{\partial z^2}-kz\left(\tilde f(z)+\left(\frac{3}{2}-\alpha\right)z\Fp\right)=0,
\label{eqfodd}
\end{equation}
for $0\leq z\leq\zs$, subject to boundary conditions  at $z=0$ and $z=\zs$.

Since $\tilde f$ is odd, we must have $\tilde f(0)=0$. The most obvious outer boundary condition, to ensure continuity of $\tilde f$, is to take $\tilde f(\zs)=0$. This way, vortensity perturbations are localised to the horseshoe region only. However, for $\xs \ll h$, we expect pressure effects to spread vortensity perturbations over a length scale $H$ \citep{casoli09}. In other words, assuming Keplerian rotation everywhere means that vortensity perturbations will not be in pressure equilibrium, while for $\xs \ll h$ we expect pressure equilibrium to be maintained at all times.

We can incorporate this in the simple model by demanding that:
\begin{equation}
\Omega^2=\okep^2 + \frac{\cs^2}{r\Sigma}\frac{\partial \Sigma}{\partial r}.
\label{eqOmega}
\end{equation}
However, using equation \ref{eqOmega} in equation \ref{eqSigma} makes the problem very difficult to solve. We can still make progress, if we assume that Keplerian rotation still holds for $x<\xs$, while we use equation \ref{eqOmega} in equation \ref{eqSigma} for $x>\xs$. In order to match the two solutions at $x=\xs$, we solve equation \ref{eqfodd} with boundary conditions $f(0)=0$, $f(\zs)=f_\mathrm{s}$, which gives us the inner solution:
\begin{eqnarray}
\tilde f_\mathrm{in}(z)=\left(\frac{3}{2}-\alpha\right)\Fp\left(\sqrt{z\zs}\frac{I_{1/3}\left(\frac{2}{3}\sqrt{kz^3}\right)}{I_{1/3}\left(\frac{2}{3}\sqrt{k\zs^3}\right)}-z\right)+\nonumber \\
f_\mathrm{s}\sqrt{\frac{z}{\zs}}\frac{I_{1/3}\left(\frac{2}{3}\sqrt{kz^3}\right)}{I_{1/3}\left(\frac{2}{3}\sqrt{k\zs^3}\right)},
\end{eqnarray}
where $I$ denotes a modified Bessel function. We fix $f_\mathrm{s}$ by demanding that the inner solution matches in a smooth way to the outer solution.  From above equation, it is clear that saturation is governed by a single parameter $p$:
\begin{equation}
p=2\sqrt{k\xs^3}/3. 
\label{eqp}
\end{equation}

\begin{figure}
\centering
\resizebox{\hsize}{!}{\includegraphics[]{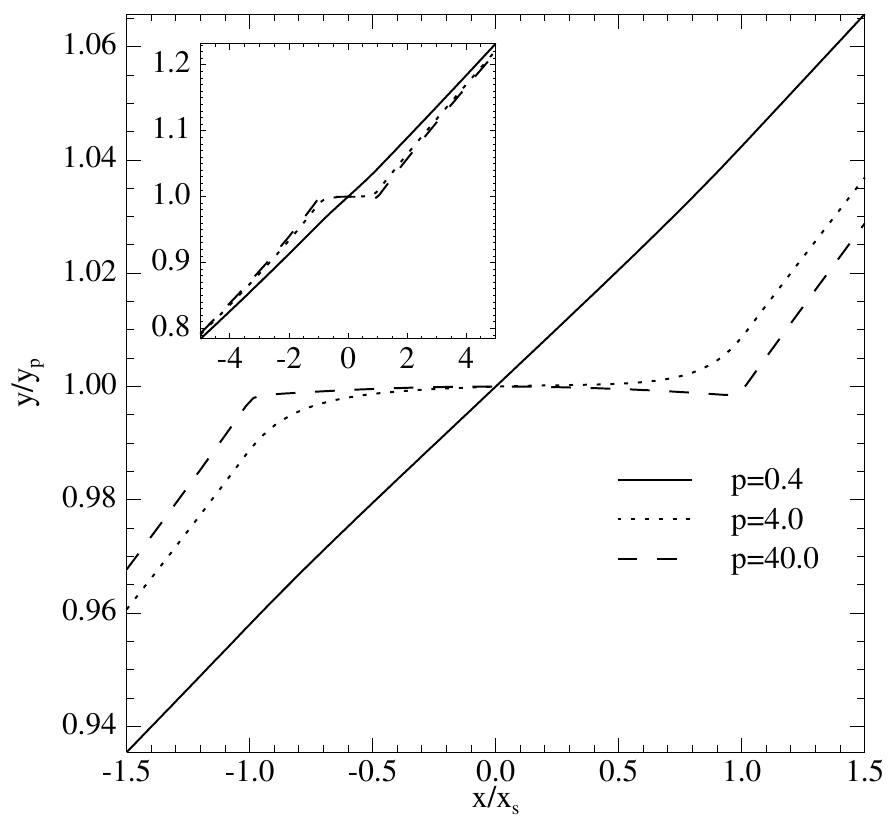}}
\caption{The function $y(x)=\Sigma(x)/\Omega(x)$, for three different saturation parameters $p$. The inset shows a larger spatial domain, where the exponential decay of the perturbation is apparent.}
\label{figyph}
\end{figure}

The outer solution $f_\mathrm{out}$ will decay on a scale $H$ towards the equilibrium solution \citep{casoli09}. There is no need to calculate it exactly, as long as we have $\xs \ll h$. We know that the odd part of the derivative of $f_\mathrm{out}$ will approach zero on a length scale $H$, so for $\xs \ll h$ we can set it to zero. We can therefore fix $f_\mathrm{s}$ by demanding that the derivative of the inner solution is zero at $\xs$. Note that this is the same boundary condition as used in \cite{mas01}. This leads to
\begin{equation}
f_\mathrm{s}=-\left(\frac{3}{2}-\alpha\right)\zs \Fp \frac{3 p I_{4/3}(p)}{2 I_{1/3}(p)+3pI_{4/3}(p)}.
\end{equation}
The resulting solution for $y(x)$ is shown in Fig. \ref{figyph}, for three different values of $p$, where we have chosen the outer profile for $x>\xs$ to approach the equilibrium profile exponentially on a scale $H$. The exact form does not matter since it does not affect the torque. Note that all profiles are indeed smooth at $x=\xs$. 

The horseshoe drag in an isothermal disc is given by
\begin{equation}
\Gamma_\mathrm{hs,baro}=3\sigp\rp^4\op^2\int_{0}^{\xs}\frac{\tilde y(x)}{\yp}x^2dx,
\end{equation}
where $\tilde y(x)$ is the odd part of $y(x)$.

\begin{figure}
\centering
\resizebox{\hsize}{!}{\includegraphics[]{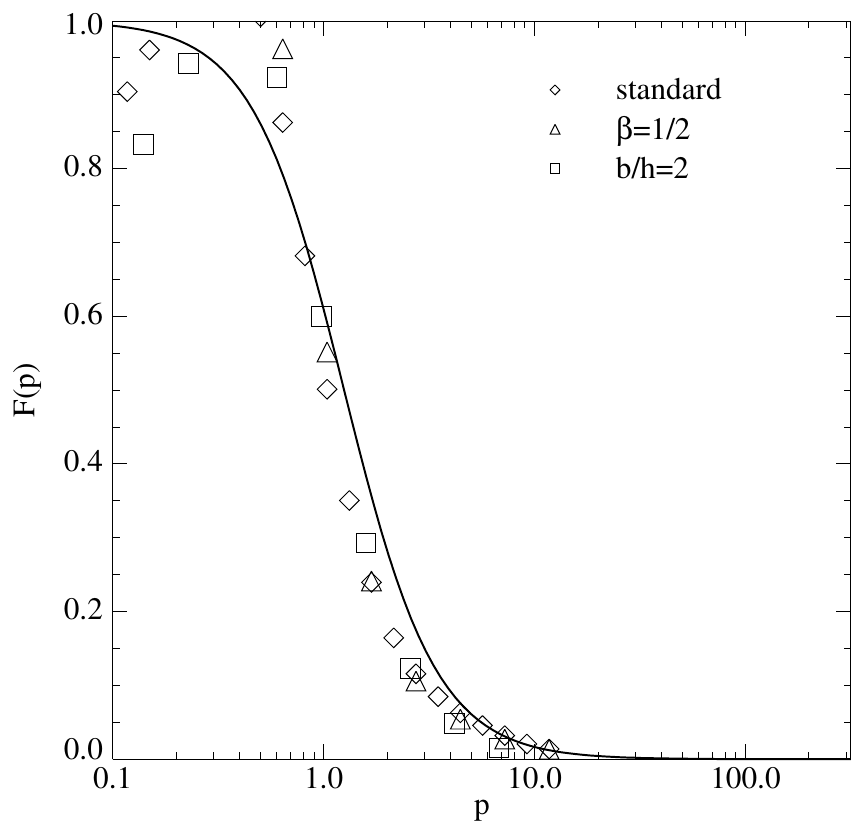}}
\caption{The function $F(p)$. The solid curve represents equation \ref{eqFph}. Symbols indicate the results of numerical simulations using {\sc fargo} of a $q=1.26 \cdot 10^{-5}$ planet embedded in a $h=0.05$ disc. The standard model has $\alpha=1/2$, $\beta=1$ and $b/h=0.4$.}
\label{figFph}
\end{figure}

The torque on the planet can be obtained by again integrating over all material that has undergone a horseshoe turn. This is why the exact form of the outer solution is of no importance: since that material never executes a horseshoe turn, it does not contribute to the torque. We then get a multiplier for the total horseshoe drag 
\begin{equation}
F(p)=\frac{8 I_{4/3}(p)}{3p I_{1/3}(p)+\frac{9}{2}p^2 I_{4/3}(p)}.
\label{eqFph}
\end{equation}
Equation \ref{eqFph} is shown in Fig. \ref{figFph}, together with numerical results obtained with {\sc fargo}. The model matches the numerical results to within $20$ $\%$ for $p>1$. For smaller values of $p$, or larger values of $\nup$, viscosity starts to affect the horseshoe turn itself, a regime we deal with in section \ref{sechighnu} below. It can be shown that equation \ref{eqFph} is identical to equation 20 in \cite{mas02}. Note, however, that in our case we automatically get the correct density profile dependence, which removes the ambiguity in the choice of $F$ in \cite{mas01}. We comment that $F(p)$ may be approximated to within $5$ $\%$ by a simpler form
\begin{equation}
F(p)=\frac{1}{1+(p/1.3)^2}.
\end{equation} 

\subsection{The limit of high $\nup$}
\label{sechighnu}
In the limit $p \ll 1$, the assumption that the horseshoe turn over time is much smaller than the viscous  diffusion time scale  across the horseshoe region breaks down. In this regime, the horseshoe turn itself is modified by viscous effects, and specific vorticity is no longer conserved  during the turn. It was shown in \cite{drag} that the non-linear horseshoe drag is then replaced by the linear corotation torque, which is in general much weaker. Because of the strong viscosity required, saturation does not play a role in this regime, making the torque fully linear. It is important to understand this regime, since in the non-barotropic case it corresponds to the transition between the optically thick, mostly adiabatic, inner disc, and the optically thin, locally isothermal, outer disc, with viscosity replaced by thermal diffusion. 

The regime in which the viscosity is strong enough to affect the horseshoe turns, but not strong enough to make the torque fully linear, is very difficult to model. All we can say is that in the limit $p \rightarrow 0$, there are basically no turns as far as the flow is concerned, since the viscosity is strong enough for the flow to forget where it came from, and the corotation torque will be fully linear. In the opposite limit, the horseshoe turns will prevent the linear torque from being set up, and the corotation torque will be given by the (possibly saturated) horseshoe drag. As the viscosity increases, the horseshoe drag decreases and the linear corotation torque increases, and it is not a-priori clear that these processes occur at the same rate. In fact, one may expect the rates to be different, since the time scale to set up the horseshoe drag, a fraction of the libration time scale, is longer than the time scale associated with the linear corotation torque, which is a dynamical time scale. 

We therefore make the assumption that the corotation torque is given by
\begin{equation}
\Gamma_\mathrm{c,baro}=G\Gamma_\mathrm{hs,baro}+(1-K)\Gamma_\mathrm{c,lin,baro},
\end{equation}
with $0 \leq G,K \leq 1$. In the limit of low viscosity (but still high enough to keep the corotation torque unsaturated), we expect $G,K\rightarrow 1$, and the corotation torque is given by the horseshoe drag. In the limit of high viscosity, $G,K \rightarrow 0$, and the torque will be fully linear. As mentioned above, we allow the horseshoe drag to decrease at a different rate than the linear torque increases towards high viscosities by having $K$ different from $G$. Below, we will derive a general form for $G$, from which we will extract $K$ using a time scale argument.

Consider the contribution of a single streamline, making a horseshoe turn from $-x$ to $x$, to the horseshoe drag. It was shown in \cite{drag} that the time scale to set up the horseshoe drag is approximately $\tturn=3\tau_\mathrm{lib}/20$. When this is comparable to the diffusion time scale across $x$, viscosity is able to affect the horseshoe drag. We take this into account by taking the contribution of the streamline to the horseshoe drag to be
\begin{equation}
d\Gamma_\mathrm{hs,baro}=3\left(\frac{3}{2}-\alpha\right)\sigp\rp^4\op^2 x^3
\mathcal{F}\left(\tau\frac{x^3}{\xs^3}\right)dx,
\end{equation}
with $\tau=\tdiff/\tturn=45\pi p^2/4$ (using $\tturn=2/(5\xs\op)$ and $\tdiff=\xs^2\rp^2/\nup$), and $\mathcal{F}(t)$ such that $\mathcal{F}(0)=0$ and $\mathcal{F}(t)\rightarrow 1$ for $t\rightarrow \infty$. For $\tturn \ll \tdiff$, this gives us the ordinary horseshoe drag, while for $\tturn \gg \tdiff$, the contribution of the streamline goes to zero. 

To keep all integrals tractable, we choose a particularly simple form for $\mathcal{F}$:
\begin{equation}
\mathcal{F}(t)=\left\{\begin{array}{rl}
(t/\tau_0)^l & t<\tau_0\\
1 & t \geq \tau_0, \end{array}\right.
\end{equation}
for some $l,\tau_0>0$. We have obtained good agreement with numerical experiments with $l=3/4$ and $\tau_0=2$. It is straightforward to show that in general
\begin{equation}
G(\tau)=4\int_0^1 w^3 \mathcal{F}(\tau w^3) dw = \frac{4}{3} \tau^{-4/3}\int_0^\tau t^{1/3}\mathcal{F}(t)dt,
\end{equation}
which for our particular form for $\mathcal{F}$ yields
\begin{equation}
G(\tau)=\left\{\begin{array}{rl}
\frac{4}{4+3l}\left(\frac{\tau}{\tau_0}\right)^l & \tau<\tau_0\\
1-\frac{3l}{4+3l}\left(\frac{\tau}{\tau_0}\right)^{-4/3}  & \tau \geq \tau_0. \end{array}\right.
\end{equation}
This can be written in terms of $p$ rather than $\tau$:
\begin{equation}
G(p)=\left\{\begin{array}{rl}
\frac{4}{4+3l}\left(\frac{45\pi}{4\tau_0}\right)^l p^{2l} & p<\sqrt{\frac{4\tau_0}{45\pi}}\\
1-\frac{3l}{4+3l}\left(\frac{4\tau_0}{45\pi}\right)^{4/3}p^{-8/3} & p \geq \sqrt{\frac{4\tau_0}{45\pi}}. \end{array}\right.
\end{equation}
For $l=3/4$, $\tau_0=2$, we finally have:
\begin{equation}
G(p)=\left\{\begin{array}{rl}
\frac{16}{25}\left(\frac{45\pi}{8}\right)^{3/4}p^{3/2} & p<\sqrt{\frac{8}{45\pi}}\\
1-\frac{9}{25}\left(\frac{8}{45\pi}\right)^{4/3}p^{-8/3}  & p\geq\sqrt{\frac{8}{45\pi}}. 
\end{array}\right.
\label{eqG}
\end{equation}

We now assume that $K$ is similar to $G$, but with a different value of $\tau_0$. Good results were obtained with $\tau_0=7$, so that the linear torque decreases slower towards lower viscosities than the horseshoe drag increases. This leads to a maximum corotation torque that is in fact larger than the unsaturated horseshoe drag, a feature that is already apparent in the numerical results shown in Fig. \ref{figFph} (see also \cite{baruteau10}). For this different value of $\tau_0$, we find
\begin{equation}
K(p)=\left\{\begin{array}{rl}
\frac{16}{25}\left(\frac{45\pi}{28}\right)^{3/4}p^{3/2} & p<\sqrt{\frac{28}{45\pi}}\\
1-\frac{9}{25}\left(\frac{28}{45\pi}\right)^{4/3}p^{-8/3}  & p\geq\sqrt{\frac{28}{45\pi}}. 
\end{array}\right.
\label{eqK}
\end{equation}

\begin{figure}
\centering
\resizebox{\hsize}{!}{\includegraphics[]{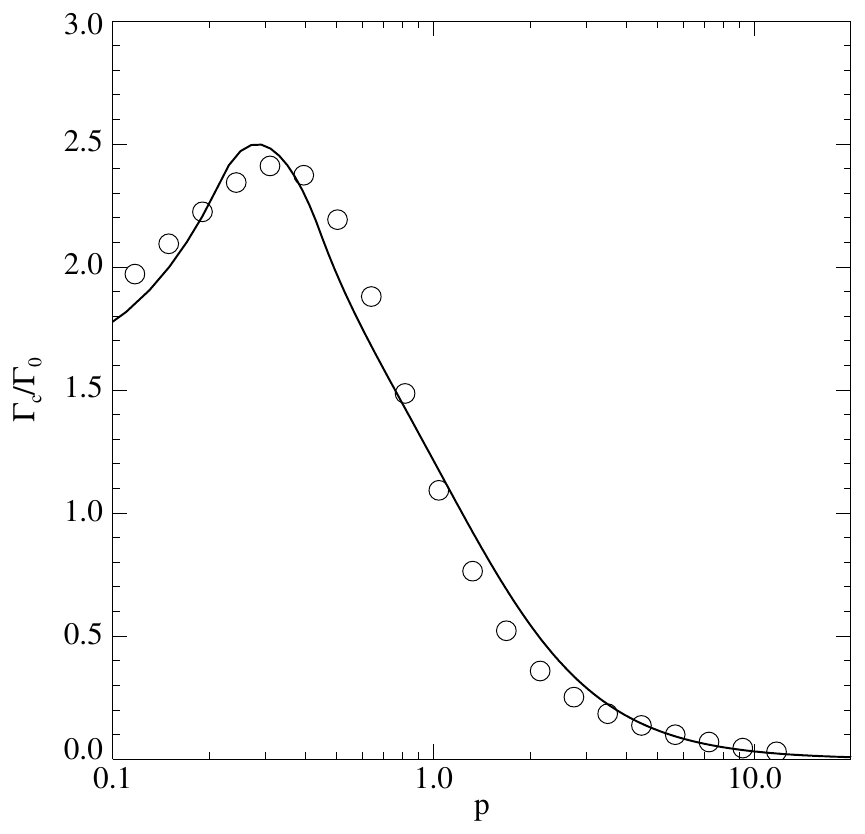}}
\caption{Corotation torque as a function of viscosity (through the $p$-parameter). Symbols indicate numerical results,  obtained with {\sc fargo}, for a $q=1.26 \cdot 10^{-5}$ planet in a locally isothermal disc with $\alpha=1/2$, $\beta=1$, $h=0.05$ and $b/h=0.4$. The solid curve denotes the model of equation \ref{eqFull}.}
\label{figisotot}
\end{figure}

We can now construct the complete model for the corotation torque, including both saturation and the cut-off at high viscosity:
\begin{equation}
\Gamma_\mathrm{c,baro}=G(p)F(p)\Gamma_\mathrm{hs,baro}+(1-K(p))\Gamma_\mathrm{c,lin,baro},
\label{eqFull}
\end{equation}
which uses equations \ref{eqFph}, \ref{eqG} and \ref{eqK}, and the torques from equations \ref{eqHSbaro} and \ref{eqCbaro}. We compare the complete model to numerical results in Fig. \ref{figisotot}, where we have used the expressions for the horseshoe drag and linear corotation torque from Paper I. As in the isothermal case, for the numerical results, $\xs$ was determined using a dichotomic search of the separatrices, with $\xs$ being calculated as the geometric average of the horseshoe half-widths at $\pm 1$ rad \citep{casoli09}. Equation \ref{eqFull} captures the main features of the numerical results quite well. We have checked that this is true for different density and temperature profiles. 

\section{Non-isothermal discs}
\label{secNonbaro}
\subsection{Disc models}
We now consider the behaviour of non-barotropic effects under the influence of diffusion. The disc models are similar to those in the isothermal case, but in addition we solve the energy equation. We stress again that we do not include viscous energy dissipation in the equations; we assume that dissipation, together with radiative cooling, sets the background temperature profile. 

\subsection{Thermal diffusion}
Since we now have an extra physical quantity that is evolved, namely the temperature, we consider thermal diffusion in addition to viscous diffusion. Since we want the disc to be in thermal equilibrium, i.e. viscous heating should compensate for radiative losses, we can constrain the range of values we need to consider for the thermal diffusion coefficient.

Radiative diffusion gives rise to a thermal conductivity
\begin{equation}
K_\mathrm{R}=\frac{acT^3}{3\kappa\rho}=\frac{4\sigma T^3}{3\kappa\rho},
\end{equation}
where $a$ is the radiation constant, $c$ the speed of light, $\sigma$ the Stefan-Boltzmann constant and $\kappa$ the opacity. The corresponding thermal diffusion coefficient $\chi$ is
\begin{equation}
\chi=\frac{K_\mathrm{R}}{c_v \rho}=\frac{4(\gamma-1)\sigma T^4}{3\kappa\rho P}=\frac{4\gamma(\gamma-1)\sigma T^4}{3\kappa\rho^2 H^2 \Omega^2},
\label{eqchi}
\end{equation}
where we have used the internal energy $\epsilon=c_v\rho T=P/(\gamma-1)$, with $c_v$ the specific heat at constant volume.

Thermal equilibrium requires that \citep{kley08}:
\begin{equation}
\sigma T_\mathrm{eff}^4=\frac{9}{4}\nu\Sigma\Omega^2,
\label{eqbalance}
\end{equation}
where $T_\mathrm{eff}$ is the effective temperature \citep{hubeny90}:
\begin{equation}
T_\mathrm{eff}^4=T^4/\tau_\mathrm{eff},
\end{equation}
with an effective optical depth
\begin{equation}
\tau_\mathrm{eff}=\frac{3\tau}{8}+\frac{\sqrt{3}}{4}+\frac{1}{4\tau},
\end{equation}
with $\tau=\kappa \rho H$ is the vertical optical depth.

Using equation \ref{eqbalance} in equation \ref{eqchi} leads to
\begin{equation}
\frac{\chi}{\nu}=\mathrm{Pr}^{-1}=\frac{9}{4}\gamma(\gamma-1)\left(1+\frac{2\sqrt{3}}{3\tau}+\frac{2}{3\tau^2}\right),
\end{equation}
where Pr is the Prandtl number, and we have taken $\Sigma=2\rho H$. We therefore have that in the limit of $\tau \rightarrow \infty$, $\mathrm{Pr}$ will be of order unity, with in general $\mathrm{Pr}<1$. In other words, thermal and viscous diffusion will be of equal magnitude in the optically thick inner regions of protoplanetary discs. In the outer regions, thermal diffusion will come to dominate, leading eventually to isothermal behaviour for $\tau \ll 1$ \citep{paard06}. Therefore, we only have to consider cases for which $\chi \geq \nu$, or, equivalently, $\mathrm{Pr }\leq 1$.  Note that this also holds in the presence of other heating sources such as external irradiation: in thermal equilibrium, any non-viscous heating source will require an increase in $\chi$ to radiate this energy away. Additional sources of heating therefore necessarily decrease $\mathrm{Pr}$ even further.

We first consider a barotropic disc, including thermal diffusion, and show that thermal diffusion does not affect the barotropic part of the corotation torque. Next, we study the case $\nup=\chip$,  with $\chip$ the thermal diffusion coefficient at the location of the planet, and show how this case is different to the barotropic case. We then adjust the model to incorporate the case $\nup \leq \chip$.

\begin{figure}
\centering
\resizebox{\hsize}{!}{\includegraphics[]{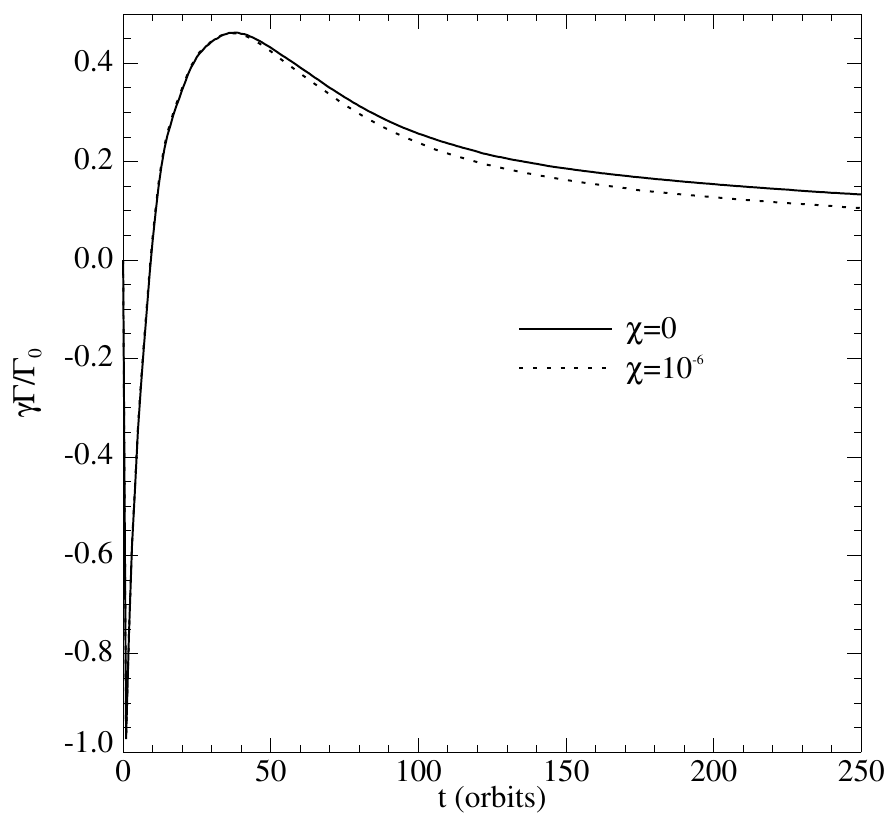}}
\caption{Total torque on a $q=1.26\cdot 10^{-5}$ planet, $b/h=0.4$, embedded in a $h=0.05$ disc with $\alpha=-1/2$, $\beta=-1/5$ and $\gamma=7/5$, making $\xi=0$. Two values for thermal diffusion are considered: $\chip=0$ (solid line), and $\chip=10^{-6}~\rp^2\op$ (dotted line).  Results were obtained with {\sc rodeo}.}
\label{figbarodiff}
\end{figure}

\subsection{Constant entropy disc}
A disc with $P=P(\Sigma)\propto \Sigma^{\gamma}$ (or constant entropy) initially behaves in a similar way to an isothermal disc. Since the entropy is constant, there is no entropy-related horseshoe drag, and since vortensity is conserved along streamlines in barotropic flows, the vortensity-related horseshoe drag is exactly the same as in the isothermal case. The only differences between the isothermal and constant entropy cases is the value of $\gamma$, which affects the total Lindblad and corotation torques (see Paper I), but also the saturation properties through its effect on $\xs$. Since $\xs$ decreases for increasing $\gamma$, for a given value of $\nup$, the horseshoe drag in a constant entropy disc with $\gamma>1$ will appear less saturated compared to the isothermal case.
 
We show in Fig. \ref{figbarodiff} that the time evolution of the torque in a constant entropy disc is not affected by thermal diffusion. Since we do not apply viscous heating or radiative cooling, entropy is conserved for $\chip=0$, keeping the disc, and the associated torque on the planet, fully barotropic. When applying thermal diffusion, entropy is no longer strictly conserved, but since both thermal and viscous diffusion act to restore the original density and temperature profiles, the disc essentially remains barotropic. The torque on the planet is therefore very similar in both cases. This is not true for very large values of $\chip$: when thermal diffusion is strong enough to affect the horseshoe turn itself, we approach the isothermal regime. This limit is considered in section \ref{seclargechi}. 

\begin{figure}
\centering
\resizebox{\hsize}{!}{\includegraphics[]{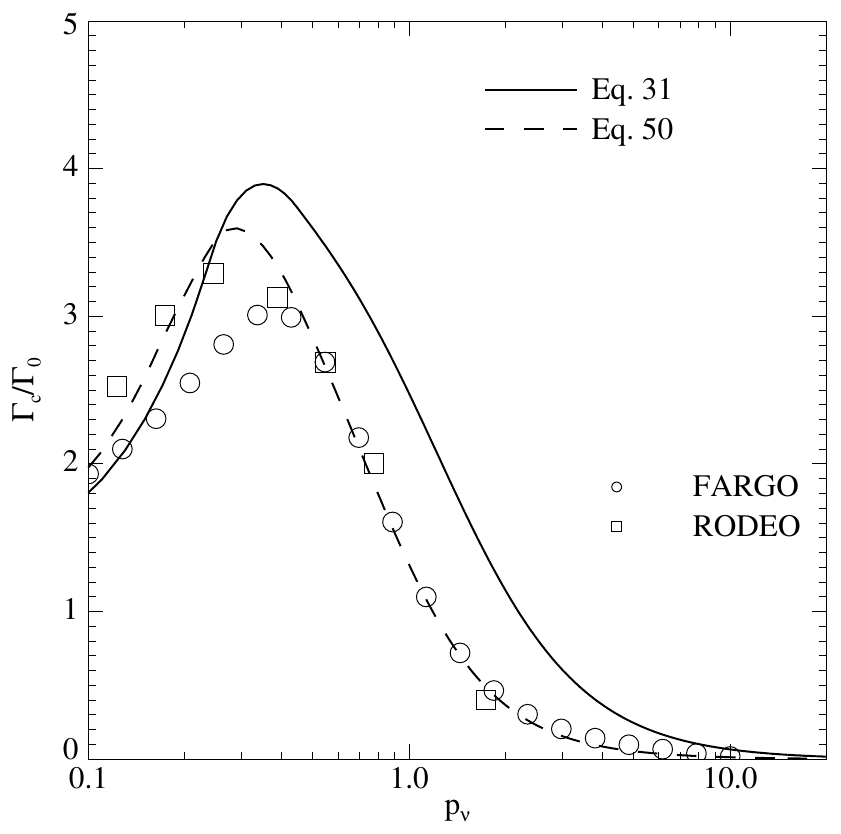}}
\caption{Corotation torque as a function of viscosity (through the $p$-parameter). Symbols indicate numerical results for a $q=1.26 \cdot 10^{-5}$ planet in a disc with $\alpha=1/2$, $\beta=1$, $h=0.05$, $\gamma=7/5$ and $b/h=0.4$. The Prandtl number equals unity in all runs. The solid curve denotes the model of equation \ref{eqFull}, with the appropriate values of the barotropic and non-barotropic linear and non-linear unsaturated torques, the dashed curve the modified model discussed in section \ref{secPr1} (see also equation \ref{eqCoroTot}).}
\label{figPr1}
\end{figure}

\subsection{The case $\nup=\chip$}
\label{secPr1}
We now consider the case $\mathrm{Pr}=1$. We have used the same disc model as for Fig. \ref{figisotot}, but now solving the energy equation with $\gamma=7/5$. Note that this disc model has negative gradients in both entropy and vortensity, making the total horseshoe drag positive, and dominated by the entropy-related part (see Paper I). The results are displayed in Fig. \ref{figPr1}. The solid curve denotes the model of equation \ref{eqFull}, with the horseshoe drag and linear corotation torque from Paper I. It is clear that the torque saturates faster than predicted by equation \ref{eqFull}. This has to do with the fact that the entropy-related part of the horseshoe drag essentially originates from a single streamline close to the separatrix. At this location, advection is at its maximum efficiency, since the velocity with respect to the planet is largest, and diffusion has a hard time to keep up. At the same time, the entropy-related part needs \emph{both} thermal and viscous diffusion to remain unsaturated. Thermal diffusion alone can not provide fresh angular momentum to the horseshoe region. Even if thermal diffusion (or entropy diffusion) can restore the entropy gradient sufficiently, if viscosity is not strong enough to diffuse the resulting vortensity production at the separatrices, the torque will still saturate \citep{paardpap08}. In this sense, the entropy torque can be doubly saturated. 

Rather than try to incorporate these effects into the simple isothermal model presented in section \ref{secSimple}, we try to adjust the model, guided by the numerical results to allow for the effects described above. Both can be incorporated by making the substitution
\begin{equation}
F(p)\rightarrow F(p_\nu)F(p_\chi),
\label{eqFnuFchi}
\end{equation} 
where $p_\chi \propto p_\nu\sqrt{\chip/\nup}$ is the saturation parameter related to thermal diffusion and $p_\nu$ is the saturation parameter associated with viscosity, in equation \ref{eqFull}. Since advection is at its most efficient in the region where the entropy-related torque is generated, we expect $p_\chi > p_\nu$ for $\nup=\chip$. Good results were obtained with $p_\chi=3p_\nu\sqrt{\chip/\nup}/2$, or
\begin{equation}
p_\chi=\sqrt{\frac{\rp^2 \op \xs^3}{2\pi\chip}},
\label{eqpchi}
\end{equation}
We can see indeed that the dashed curve in Fig. \ref{figPr1}, obtained with equation \ref{eqFnuFchi}, gives a good fit to the data at large $p_\nu$. Note that the substitution equation \ref{eqFnuFchi} needs to be made for the entropy-related part of the horseshoe drag only.

The cut-off at high $\nu$ and $\chi$ should also be modified. Note that in this regime, advection being at its most efficient actually helps to keep the torque non-linear at higher levels of diffusion. Rather than defining new functions $G$ and $K$, we have found that the numerical results can be fitted by making the substitutions
\begin{equation}
G(p)\rightarrow \sqrt{G(p_\nu)G(p_\chi)}
\end{equation}
and
\begin{equation}
1-K(p)\rightarrow \sqrt{(1-K(p_\nu))(1-K(p_\chi))}.
\end{equation}
Here again, these substitutions only concern the entropy-related part of the horseshoe drag. The resulting model is shown by the dashed curve in Fig. \ref{figPr1}. Note that at higher levels of diffusion, the {\sc rodeo} results differ by 20\% from results obtained with {\sc fargo}. We have verified that half of this is due to the different diffusion implementation: entropy diffusion implemented in {\sc rodeo} gave lower values for the torque by approximately 10\% at high diffusion. The remaining 10\% can be attributed to a higher value of the unsaturated torque using {\sc rodeo} (see Paper I). This dependence on the nature of the diffusion should be kept in mind when interpreting these results in terms of radiative discs, where $\chi$ is due to radiative diffusion.  The error in the model compared to the simulations can thus be up to $20$ $\%$.

\begin{figure*}
\centering
\resizebox{\hsize}{!}{\includegraphics[]{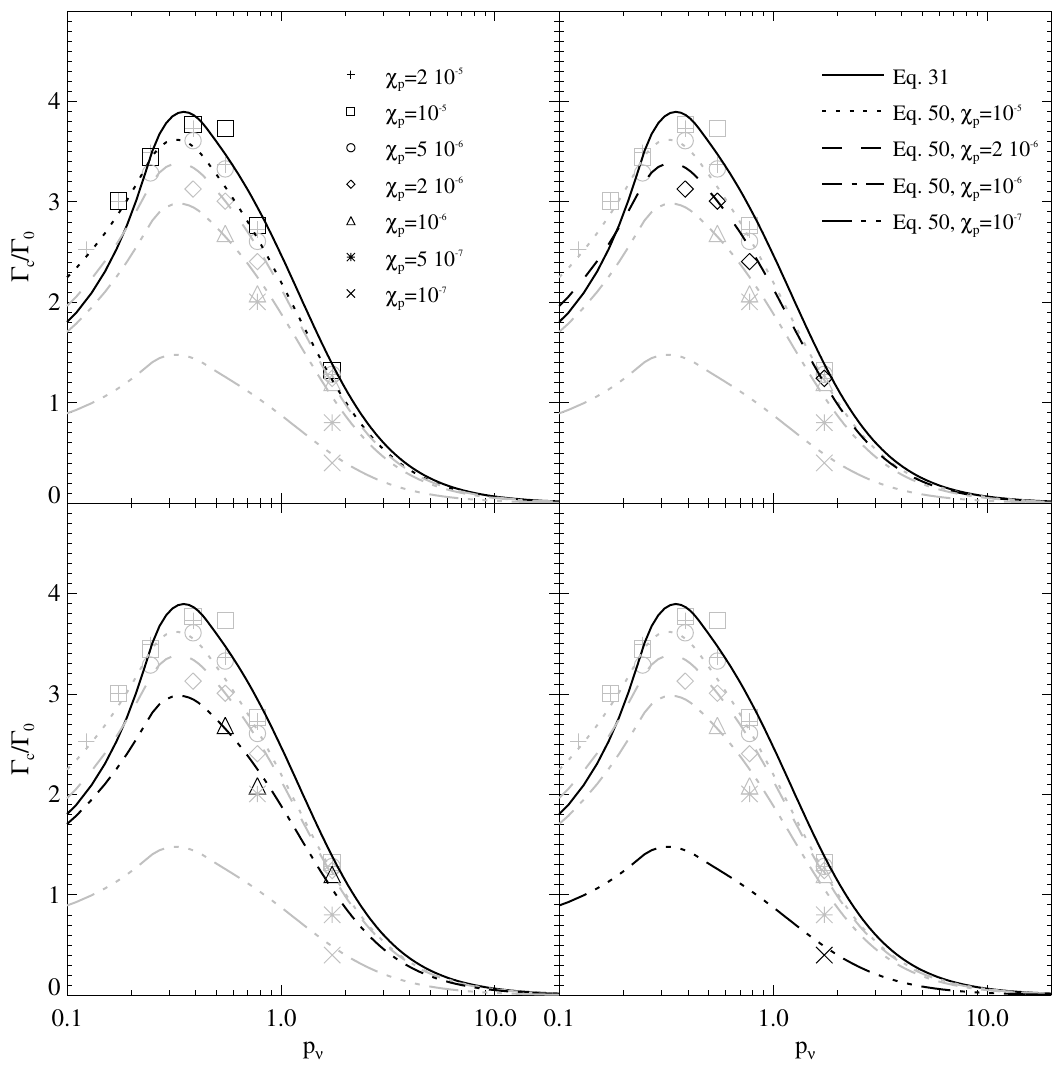}}
\caption{Corotation torque as a function of viscosity (through the $p$-parameter). Symbols indicate numerical results for a $q=1.26 \cdot 10^{-5}$ planet in a disc with $\alpha=1/2$, $\beta=1$, $h=0.05$, $\gamma=7/5$ and $b/h=0.4$, for different values of thermal diffusion $\chip$. The solid curve denotes the model of equation \ref{eqFull}  (saturation of the non-barotropic horseshoe drag is similar to the barotropic part), with the appropriate values for the barotropic and non-barotropic parts of the linear and non-linear unsaturated torques. The remaining curves indicate the modified model discussed in section \ref{secPr1} for different values of $\chip$ (see also equation \ref{eqCoroTot}).  Different panels highlight different values of $\chip$, with the corresponding model prediction. Numerical results were obtained with {\sc rodeo}.}
\label{figenetot}
\end{figure*}

\subsection{The general case $\nup \leq \chip$}
\label{seclargechi}
We now consider the general case, still with the restriction $\mathrm{Pr} \leq 1$. We have run a grid of models with different $\nup$ an $\chip$, and the results are displayed in Fig. \ref{figenetot}. All numerical results were obtained with {\sc rodeo}, using thermal diffusion. 

For all values of $p_\nu$, $\mathrm{Pr}=1$ gives the lowest value for the corotation torque. Going up in thermal diffusion, at a fixed value of $p_\nu$, the torque increases towards the original model, denoted by the solid curve. This is consistent with the effect discussed in the previous section, that a higher thermal diffusion coefficient is needed to fight advection of entropy along the separatrix.  For all values of $\chip$, the improved model gives a reasonable fit to the numerical results,  with all deviations less than approximately $20$ $\%$.

\begin{figure}
\centering
\resizebox{\hsize}{!}{\includegraphics[]{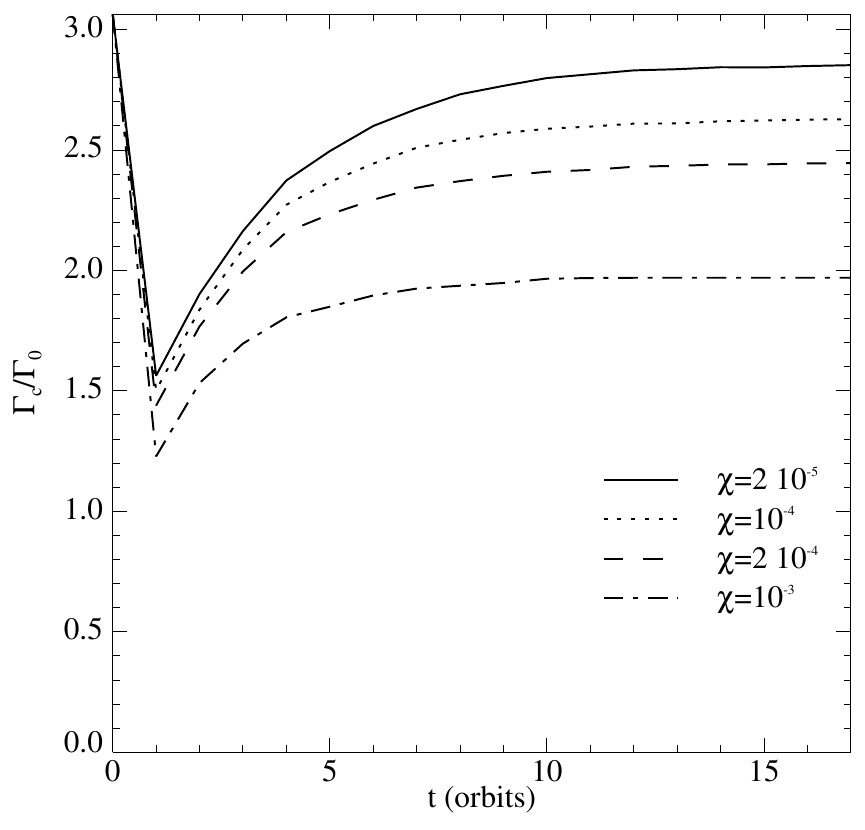}}
\caption{Corotation torque on a $q=1.26 \cdot 10^{-5}$ planet in a disc with $\alpha=1/2$, $\beta=1$, $h=0.05$, $\gamma=7/5$ and $b/h=0.4$, $\nup=10^{-5}$ for different values of thermal diffusion $\chip$.  Results were obtained with {\sc rodeo}.}
\label{figlargechi}
\end{figure}

For $\chip >10^{-5}$, the corotation torque starts to decrease again. This is due to thermal diffusion starting to affect the horseshoe turns. For $\chip$ large enough, we expect to enter the locally isothermal regime, where the corotation torque is given by the barotropic, non-linear horseshoe drag, plus the linear, entropy-related corotation torque. Since the linear corotation torque is in general smaller than its non-linear counterpart, the corotation torque will decrease towards higher values of $\chip$. This is illustrated in Fig. \ref{figlargechi}, where we show the time evolution of the corotation torque at a fixed value of $\nup=10^{-5}$ for different values of $\chip$. The torque decreases with $\chip$, but the barotropic part of the corotation torque remains non linear. The barotropic horseshoe drag plus the linear entropy-related corotation torque amount to $1.4\Gamma_0$. However, at these high values of $\chip$, we also need to consider the effect of thermal diffusion on the effective value of $\gamma$, which will approach unity for large values of $\chip$. This we discuss next. 

\begin{figure}
\centering
\resizebox{\hsize}{!}{\includegraphics[]{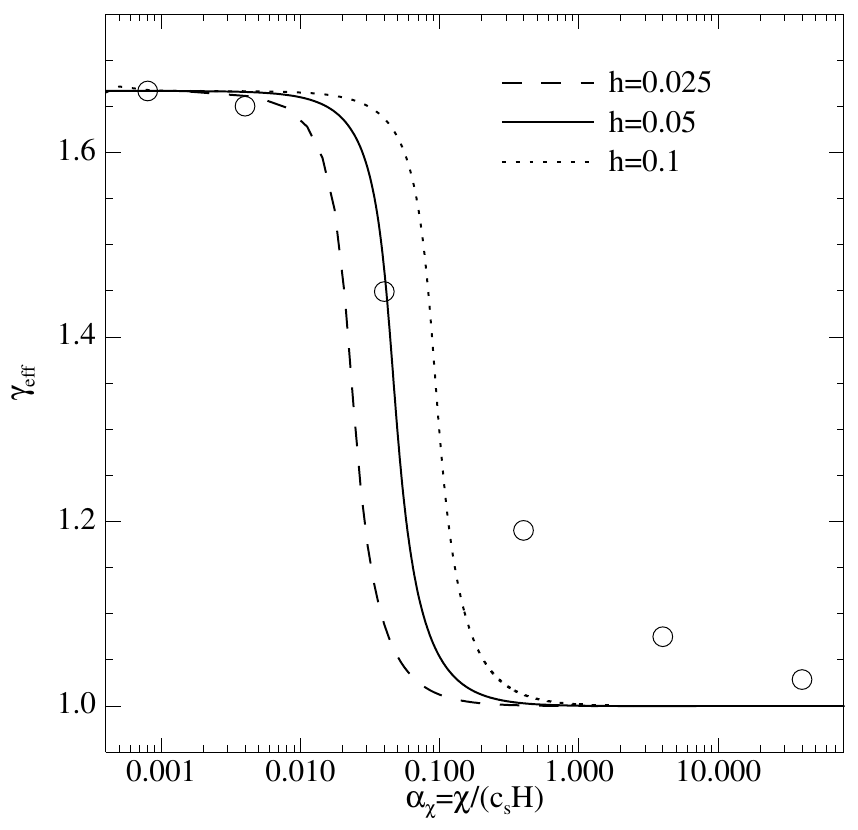}}
\caption{The values of $\gamma_\mathrm{eff}$ from equation \ref{eqgammaeff}, as a function of thermal diffusion, here represented by $\alpha_\chi = \chip/(\cs\op)$.  Curves denote equation \ref{eqgammaeff} for different values of $h$; open circles indicate numerical results obtained with {\sc rodeo} for $h=0.05$.}
\label{figgammaeff}
\end{figure}

\subsection{Lindblad torque}
The Lindblad torque is due to a superposition of linear waves, and does not strongly depend on viscosity, for reasonable values of the viscosity $\nu$. Thermal diffusion, on the other hand, can play an important role, and is necessary to recover the isothermal result for large values of the thermal diffusivity $\chip$. The difference between the isothermal and adiabatic Lindblad torque basically results from a difference in sound speeds. Since $\Gamma_\mathrm{L}\propto \cs^{-2}$, $\Gamma_\mathrm{L,iso}/\Gamma_\mathrm{L,adi}=\gamma$. 

Consider the dispersion relation for plane waves $\propto \exp(i(\omega t-kx))$ \citep{mihalas84}:
\begin{equation}
\omega^2=\cs^2k^2 \frac{1-i\chi k^2/\omega}{1-i\gamma\chi k^2/\omega},
\end{equation}
with $\cs$ the adiabatic sound speed. From this, one can derive a phase speed $v_\mathrm{p}$: 
\begin{eqnarray}
\lefteqn{\frac{v_\mathrm{p}^2}{\cs^2}=}\nonumber \\
\lefteqn{\frac{2Q}{\gamma Q+\frac{1}{2}\sqrt{2\sqrt{(\gamma^2Q^2-1)^2+(4Q-2\gamma Q)^2}+2\gamma^2Q^2-2}},}
\end{eqnarray}
with $Q=\chi\omega/\cs^2$. In the limit of no thermal diffusion ($Q=0$), $v_\mathrm{p}\rightarrow \cs$: waves propagate at the adiabatic sound speed. For $Q\rightarrow\infty$, $v_\mathrm{p}\rightarrow \cs/\gamma$: waves propagate at the isothermal sound speed.

The torque for any Fourier component will now scale as $v_\mathrm{p}^{-2}$, which makes the isothermal torque a factor of $\gamma$ stronger than the adiabatic torque. In principle, $\omega=m\op$ for the problem at hand, and one should therefore consider the effects of diffusion on each Fourier component separately. For simplicity, we consider a single, representative value of $m$ to describe the effects of diffusion. Since most of the torque comes from $m \sim 2/3h$, we set
\begin{equation}
Q=\frac{2\chip\op}{3h\cs^2}=\frac{2\chip}{3h^3\rp^2\op}.
\end{equation}
It is convenient to define an \emph{effective} $\gamma$:
\begin{eqnarray}
\lefteqn{\gamma_\mathrm{eff}=}\nonumber \\
\lefteqn{\frac{2Q\gamma}{\gamma Q+\frac{1}{2}\sqrt{2\sqrt{(\gamma^2Q^2+1)^2-16Q^2(\gamma-1)}+2\gamma^2Q^2-2}},}
\label{eqgammaeff}
\end{eqnarray}
so that
\begin{equation}
\gamma_\mathrm{eff} \Gamma_\mathrm{L}/\Gamma_0=-2.5-1.7\beta+0.1\alpha,
\label{eqLindeff}
\end{equation}
which then replaces equation \ref{eqLind}. Note that the horseshoe drag also scales with $\gamma_\mathrm{eff}$, since the horseshoe width depends on $\gamma$ \citep{horse}. Equation \ref{eqgammaeff} is plotted in Fig. \ref{figgammaeff} as a function of $\alpha_\chi = \chip/(\cs H)$, for three different scale heights $h$. It is clear that a substantial thermal diffusion coefficient is required in order to modify the Lindblad torque (typically $\chip=10^{-3}\rp^2\op$). We have overplotted numerical results obtained with {\sc rodeo}, for a disc with $h=0.05$, $\alpha=3/2$, $\beta=1$ and $\gamma=5/3$. Since this disc has no entropy or vortensity gradients, the corotation torque vanishes.  We can therefore measure $\gamma_\mathrm{eff}$ from the total torque, which should be given by equation \ref{eqLind} with $\gamma=\gamma_\mathrm{eff}$. Since for equation \ref{eqgammaeff} we have chosen a single value of $m$ to signal the transition from isothermal to adiabatic, we expect the numerical results to show a smoother transition between $\gamma_\mathrm{eff}=5/3$ and $\gamma_\mathrm{eff}=1$. This is indeed what is observed in Fig. \ref{figgammaeff}. Note, however, that the model does a good job in predicting where the transition takes place.  We have checked that this also holds for different values of $h$. Around this transition, the maximum error in $\gamma_\mathrm{eff}$ observed is $20$ $\%$. Note that, since all unsaturated torques are proportional to $\gamma_\mathrm{eff}^{-1}$ (see below), the error in $\gamma_\mathrm{eff}$ only affects the magnitude of the total torque, not the sign. The dependence of $\gamma_\mathrm{eff}$ on thermal diffusion will therefore not affect planet migration in a qualitative way, but it is necessary to connect the adiabatic to the isothermal regime.

\subsection{A modified torque formula}
\label{secForm}
Before we combine the results of the previous sections into a single torque formula,   we restate the assumptions made in deriving these results. We have worked in the two-dimensional approximation, considered laminar discs only, and kept the planet on a fixed circular orbit. We have ignored the self-gravity of the disc, assumed thermal equilibrium,  and considered a simplified energy equation. It is clear that future studies should be aimed at clarifying the role of magnetic turbulence and radiative energy transport in keeping the corotation torque unsaturated.

We have mainly focused on a $q=1.26\cdot 10^{-5}$ using a softening parameter $b/h=0.4$, which seems to be an appropriate value to mimic 3D effects \citep{drag}. In the model, the only place where the mass and the softening parameter enter is in the values of the unsaturated torques, which have a $q^2$ dependence and depend on $b/h$ \citep{drag}, and in the value of $\xs$, which can be written as \citep{horse}
\begin{equation}
\xs=C(b/h) \sqrt{\frac{q}{h}},
\label{eqxs}
\end{equation}
where $C$ can be approximated by a power law around $b/h=0.4$:
\begin{equation}
C=\frac{1.1}{\gamma_\mathrm{eff}^{1/4}}\left(\frac{0.4}{b/h}\right)^{-1/4},
\label{eqC}
\end{equation}
(see Paper I). The dependence on softening of the linear torques was studied in \cite{drag}, and of $\xs$ and the non-barotropic horseshoe drag in Paper I. Given a different value of the softening, it is straightforward to adjust the formulae below by using equations \ref{eqxs} and \ref{eqC}. The value of $\xs$ then features in the saturation parameters $p_\nu$ (given by equation \ref{eqp}) and $p_\chi$ (equation \ref{eqpchi}). Note in particular that the functional form of $F$, $G$ and $K$ will not change for different values of $b/h$.

The total torque on a low-mass planet is given by the Lindblad torque plus the corotation torque:
\begin{equation}
\Gamma=\Gamma_\mathrm{L}+\Gamma_\mathrm{c},
\end{equation}
with $\Gamma_\mathrm{L}$ given by equation \ref{eqLind}, but with $\gamma\rightarrow \gamma_\mathrm{eff}$ (see equation \ref{eqLindeff}).

The corotation torque is split into a barotropic part and an entropy-related part:
\begin{equation}
\Gamma_\mathrm{c}=\Gamma_\mathrm{c,baro}+\Gamma_\mathrm{c,ent}.
\label{eqCoroTot}
\end{equation}
The barotropic part of the horseshoe drag is not affected by thermal diffusion, therefore
\begin{equation}
\Gamma_\mathrm{c,baro}=\Gamma_\mathrm{hs,baro}F(p_\nu)G(p_\nu)+(1-K(p_\nu))\Gamma_\mathrm{c,lin,baro} 
\end{equation}
where $\Gamma_\mathrm{hs,baro}$ is the fully unsaturated horseshoe drag (equation \ref{eqHSbaro}, but with $\gamma\rightarrow \gamma_\mathrm{eff}$), $\Gamma_\mathrm{c,lin,baro}$ is the linear barotropic corotation torque (equation \ref{eqCbaro}, but with $\gamma\rightarrow \gamma_\mathrm{eff}$), $F(p)$ governs saturation (see equation \ref{eqFph}), and $G(p)$ and $K(p)$ govern the cut-off at high viscosity (see equations \ref{eqG} and \ref{eqK}). 

For the non-barotropic, entropy-related corotation torque, we need the modifications discussed in section \ref{secPr1}:
\begin{eqnarray}
\Gamma_\mathrm{c,ent}=\Gamma_\mathrm{hs,ent}F(p_\nu)F(p_\chi)\sqrt{G(p_\nu)G(p_\chi)}+\nonumber \\
\sqrt{(1-K(p_\nu))(1-K(p_\chi))}\Gamma_\mathrm{c,lin,ent},
\end{eqnarray}
where $\Gamma_\mathrm{hs,ent}$ is the fully unsaturated, entropy-related part of the horseshoe drag (equation \ref{eqHSent}, but with $\gamma\rightarrow \gamma_\mathrm{eff}$), $\Gamma_\mathrm{c,lin,ent}$ is the linear, entropy-related part of the corotation torque (equation \ref{eqCent}, but with $\gamma\rightarrow \gamma_\mathrm{eff}$), and $p_\chi$ is the saturation parameter associated with thermal diffusion (see equation \ref{eqpchi}).

From Fig. \ref{figenetot}, we estimate that the maximum error in the corotation torque is approximately $20$ $\%$ compared to the simulations. From Paper I, we know that the error in the Lindblad torque is smaller in general, so for the total torque we expect differences of $20$ $\%$ between model and simulations as well, except close to the transition between isothermal and adiabatic behaviour, where the error in the Lindblad torque can be $20$ $\%$ as well (see Fig. \ref{figgammaeff}).

\section{Discussion}
\label{secDisc}
We have presented numerical results and some simple models that capture the behaviour of the torque due to Lindblad resonances and horseshoe drag on low-mass planets embedded in gaseous discs in the presence of viscous and thermal diffusion. Several simplifications of the real physical system were made to keep the problem tractable.

First of all, we have worked with vertical integrated quantities only, which makes the problem two-dimensional. The entropy-related horseshoe drag seems to play an important role in three-dimensional discs as well \citep{kley09}, and it has been shown that, in an isothermal set-up, the effect of the horseshoe drag is stronger in three dimensions \citep{masset06}. Clearly, a three-dimensional model of the horseshoe region is desirable. This would also get rid of the dependence of the torque on the softening parameter that is necessarily part of a two-dimensional model (see Paper I). The results of \cite{kley09} indicate, however, that the main features of three-dimensional simulations can be reproduced nicely within the two-dimensional approximation.

We have only considered laminar, viscous discs. In reality, discs are expected to be turbulent in regions where the gas couples to the magnetic field of the disc \citep{balbus}. The interaction of turbulence with horseshoe turns is an area that is still largely unexplored. For the isothermal case, with a simplified turbulence model based on stochastic forcing, \cite{baruteau10} showed that turbulent models show similar behaviour as laminar models. It remains to be seen how full magnetic turbulence, in a non-isothermal setting, affects the horseshoe drag.

We have modelled radiative cooling as thermal diffusion, with a constant thermal diffusion coefficient. In radiative models, $\chi$ will depend strongly on temperature and opacity. Moreover, radiative diffusion will probably be most efficient in the vertical direction, while in our two-dimensional approximation thermal diffusion is restricted to the plane of the orbit of the planet. It was noted in \cite{kley08} that this is in fact an unfavourable case for sustaining the corotation torque compared to vertical cooling in radiative models.  Note that we have chosen $\chi$ such that the thermal diffusion time scale is identical to the vertical cooling time scale. Since both act to restore the original temperature profile, it is to be expected that including heating and cooling (without thermal diffusion) will lead to similar effects as thermal diffusion alone. When acting together, they may reduce the effective Prandtl number since the flow then has two ways of restoring the original temperature profile.

We have taken the background gradients of temperature and surface density to be given. In reality, they will be determined by a balance of radiative heating and cooling and viscous dissipation. It was shown recently in \cite{lyra} that discs where these profiles are calculated self-consistently allow for outward Type I migration to occur for a large fraction of the life time of the disc. Of course, since the mass of the disc decreases with time, radiative cooling becomes more and more efficient, leading in the end to a situation where the entropy-related torques are fully linear. 
 
The strong dependence of the horseshoe drag on the local temperature gradient makes it relatively easy to trap planets at special locations in the disc. While in the isothermal case, a surface density jump \citep{trap} or at least a positive surface density gradient \citep{drag} is required, in the non-isothermal case a trap is readily established if a local negative temperature gradient exists. Even if the disc as a whole does not permit outward Type I migration, low-mass planets may stop their inward journey when reaching such a temperature trap. 

If the disc parameters allow for outward migration, a natural planet trap occurs at the transition between the optically thick and optically thin regions of the disc. Outside this radius, we have isothermal, and therefore inward, Type I migration, while inside this radius, we have adiabatic, and therefore outward, Type I migration. The evolution of this equilibrium radius was explored in more detail in \cite{lyra}. Because of the dependence of the corotation torque on the saturation parameter $p$, which depends on the planet mass through $\xs$, different planet masses will have different equilibrium radii. The impact of this migration behaviour on the gravitational interactions between planets are still largely unexplored. 

We have focused on low-mass planets, comparable to the Earth. While planets of approximately 1 Jupiter mass start to open up deep annular gaps in the disc \citep{linpap86}, there exists an interesting intermediate regime where planets comparable to Neptune experience an enhanced horseshoe drag \citep{masset06}. This was interpreted in \cite{horse} as being due to the ineffectiveness of the Lindblad wake to push against horseshoe region, leading to a larger value of the horseshoe width at these masses. This means that these planets are more readily subject to outward migration, leading to mass segregation even without considering saturation.  

 The planet was kept on a fixed circular orbit throughout this paper, and we have ignored any radial viscous flow in the gas. If the relative radial speed of the planet with respect to the gas is large enough, the streamline topology will be modified, leading to a flow structure similar to that encountered in Type III migration \citep{maspap03,adam08}. Although in this case the discussion on the corotation torque will need to be modified, we comment that a very massive disc, with a Toomre $Q$ of order unity, is needed for a low-mass planet to migrate fast enough to achieve this situation.

Finally, we have ignored the self-gravity of the disc. The impact of self-gravity on Type I migration was discussed in \cite{pierens} and \cite{selfgrav}, finding an acceleration of inward migration due to a shift in the Lindblad resonances. This should be taken into account especially in massive discs. A quantitative estimate of this effect can be inferred from Fig. 8 of \cite{selfgrav}.

\section{Conclusions}
\label{secCon}
We have presented a simple model that can capture the complicated behaviour of the corotation torque with respect to viscous and thermal diffusion. The formulae presented in section \ref{secForm} show good agreement with two-dimensional hydrodynamic simulations  to within $20$ $\%$. The strong dependence of the total torque on temperature gradients and the level of viscous and thermal diffusion allow for a wealth of possible outcomes of Type I migration, depending on global disc parameters.

\section*{Acknowledgements}
We thanks the referee for an insightful report that helped us improve the clarity of the paper. It is a pleasure to thank A. Crida and K. Kretke for a careful reading of the manuscript. SJP acknowledges support from STFC in the form of a postdoctoral fellowship. Simulations using {\sc rodeo} were performed using the Darwin Supercomputer of the University of Cambridge High Performance Computing Service (http://www.hpc.cam.ac.uk), provided by Dell Inc. using Strategic Research Infrastructure Funding from the Higher Education Funding Council for England. {\sc fargo} simulations were performed using the Pleiades cluster at UC Santa Cruz.

\bibliography{paardekooper.bib}

\label{lastpage}

\end{document}